\documentclass[a4paper]{jpconf}
\usepackage{graphicx}
\usepackage{epsfig,amssymb,multirow,amsmath, bm} 
\usepackage{tikz}

\begin{document}
	\title{The lambda hyperon and the hyperon puzzle}
	
	\author{Wazha German$^{1}$ and Jacobus Diener$^{1,2}$}
	
	\address{$^1$Department of Physics and Astronomy, Private Bag 16, Botswana International University of Science and Technology, Palapye, Botswana}
	\address{$^{2}$Centre for Higher and Adult Education, Faculty of Education, Stellenbosch University, Matieland, South Africa}
	
	\ead{wazha.german@studentmail.biust.ac.bw}
	
	\begin{abstract}
		Neutron stars provide unique conditions to study cold dense nuclear matter at extreme densities. Due to these extreme conditions additional hadronic degrees of freedom are expected to be populated,including hyperons. This talk will focus on the influence of hyperons on the neutron star equation ofstate. In particular the contribution of the lambda hyperon will be discussed, as a first approximation to describing exotic neutron star equations of state. The system under consideration is where the strong nuclear force is described by the exchange of mesons and applying the relativistic mean field theory to study dense nuclear matter. As expected, the inclusion of the lambda hyperon softens the neutron star equation of state (EoS). A softer EoS will reduce the maximum mass attainable by the modeled neutron star with such EoS. While hyperons are certainly not unexpected in high density systems, but there presence seems to be contradicted by observations of high mass neutron stars.This contradiction is known as the “hyperon puzzle”. The expected influx of observational data from massive new radio-telescopes like the Square Kilometer Array (SKA) will provide observations that can be supported and evolve theoretical models of nuclear matter. Therefore, the study of hyperonic matter is not only relevant to nuclear theory, but also locally to Botswana as an African partner country of the SKA.
	\end{abstract}
	
	\section{Introduction}
	Neutron stars are remnants formed from the gravitational core collapse of giant stars during a type II supernovae. They are made up of dense nuclear matter $10^{17}$ kg/m$^3$
	, a radius in the range of $10$ - $12$ km and masses around $1$-$2\ M_{\odot}$ (solar masses) \cite{Serot}. In addition, the densities in neutron stars are in the range of $4$ - $8$ times the nuclear saturation density ($\rho_{sat} = 0.153 $ fm$^{-3}$) making them one of the most dense objects in the universe \cite{Glendenning}. Due to such conditions neutron stars provide ideal laboratories to study nuclear matter at high densities as such conditions cannot be replicated in terrestrial labs \cite{Glendenning}.\\
	\\
	The equation of state (EoS) is a relation between the pressure and energy density in a fluid. Applied to modelling a neutron star, the EoS will describe the properties of the matter in the interior of a neutron star \cite{Prakash}. 
	Due to be being small and very far away so it is difficult to observe neutron stars and correspondingly to probe the EoS directly. Therefore neutron star modelling relies heavily on models for nuclear matter, due to the similarities in density. \\
	\\
	A typical neutron star is modelled as a strong interacting system of matter composed of nucleons and leptons at zero temperature. The equilibrated neutron star is assumed as charge neutral, otherwise 
	the Coulomb force will overcome the gravity that bounds the star and will rip it apart, as well as in beta equilibrium \cite{Serot}. Due to extreme neutron star densities, additional hadronic degrees of freedom are expected to be populated \cite{Haensel}. These additional hadronic degrees of freedom includes hyperons (baryons with strange quark content) since the equilibrium conditions in neutron stars can make the capture of hyperons energitically favourable \cite{Vidana}. The presence of hyperons (or any additional degrees of freedom for that matter) in the neutron star leads to a softening of the EoS (less rapid increase in the pressure of the system with increasing density) \cite{Logoeta}.  A softer neutron star EoS leads to the reduction in maximum neutron star mass that can be supported by the EoS (the neutron star mass-radius relationship for a specific EoS is obtained by solving the general relativistic equations for the hydrostatic equilibrium, the TOV equations \cite{Glendenning}).\\
	\\
	Since higher mass neutron star will have a higher densities in the star's interior, these stars would be expected to exhibit the most high energy behaviour.  However, the inclusion of more (higher energy) particle degrees of leads to softening of the EoS and thus the reduction of the maximum mass.	This is known as the ``hyperon puzzle''. This creates a "puzzle" because the effect hyperons have on neutron stars contradicts maximum masses from recent observations of pulsars PSR J1614-2230 $(1.97 \pm 0.04) M_{\odot}$ \cite{Demorest} and PSR J0348+0432 $(2.01\pm 0.04)M_{\odot}$ \cite{Antoniadis}. This leads to maximum masses that are not compatible with observations \cite{Bombaci} since these high mass neutron star observations rule out nearly all hyperon models for the EoS.\\
	\\
	Our contribution will focus on the investigating the hyperon puzzle using a relativistic mean field (RMF) for nuclear matter that includes the lambda hyperons (lightest hyperon) as a first approximation, based on W German's MSc study. 
	\section{Methodology}
	RMF is a relativistic description of nuclear matter where the strong nuclear force is described by the exchange of mesons \cite{Serot, Glendenning}. 
	\begin{eqnarray}
		{\cal L} &=&{\cal L}_{\mbox{{\footnotesize Dirac}}}+{\cal L}_{\mbox{{\footnotesize KG}}}+{\cal L}_{\mbox{{\footnotesize Proca}}}
		-g_{v}\bar{\psi}_D\gamma_{\mu}\omega^{\mu}\psi_D
		+g_{s}\bar{\psi}_D\phi\psi_D 
		- \frac{g_\rho}{2}\bar{\psi}_N{\bm\tau}\cdot{\bm b}_\mu\psi_N
		\nonumber\\
		&& - \frac{\kappa}{3!}\big(g_s\phi\big)^3 - \frac{\lambda}{4!}\big(g_s\phi\big)^4.\label{Lgrngn}
	\end{eqnarray}
	Here ${\cal L}_{\mbox{{\footnotesize Dirac}}}$ is the free field component of the Lagrangian for the neutrons, protons and lambda hyperons, ${\cal L}_{\mbox{{\footnotesize KG}}}$ is the Klein-Gordon Lagrangian for the scalar sigma ($\sigma$) mesons and likewise ${\cal L}_{\mbox{{\footnotesize Proca}}}$ describes massive omega vector ($\omega$) and isovector rho ($\bm b$) mesons. $\psi_D$ describes the general Dirac field spinor, while $\psi_N$ is only the nucleon (protons and neutrons) Dirac field spinor\footnote{Since the lambda hyperon has isospin $0$, it does not couple to the isovector rho meson.}.	For the nucleon-meson coupling strengths $g_v$, $g_s$, and $g_\rho$ the values of the QHD1 \cite{Serot} and NL3 \cite{NL3} parameter sets were used. The coupling constants are fixed so that various properties of nuclei and nuclear matter are consistently described \cite{Serot}.  The $\kappa$- and $\lambda$-couplings are addition self-couplings of the scalar meson field that are introduced in the NL3 parameter pertaining to nucleon reduced mass to improve the description of the nuclear compressibility \cite{NL3}.\\
	\\	
	The scalar meson fields from the scalar sigma meson give rise to a strong attractive central force whilst the omega vector meson gives rise to a strong repulsive central force.  The rho meson accounts for the isospin interaction of the nucleons. The EoS based on Eq.\ \ref{Lgrngn} is calculated using the RMF approximation, where the meson fields are replaced by the their ground state expectation values \cite{Serot}.
	As such the RMF is an approximation method that becomes increasingly valid at higher densities. 
	Unfortunately, unlike Quantum Electrodynamics (QED) where a pertubative approach can be applied and all the terms can solved exactly, in the RMF the higher order terms diverge (due to the coupling strengths being large) and thus a pertubation expansion cannot be applied in these nuclear matter descriptions \cite{Lattimer}.\\
	\\
	The EoS for 
	pure neutron matter, a beta equilibrated, charge neutral model and finally a beta equilibrated charge neutral system with the lambda hyperon states populated and the resulting neutron star properties (mass-radius relationships) compared. \\
	\\
	To get the mass-radius relationship, the EoS is applied to the Tolman-Oppenheimer-Volkoff equation (TOV equation). The TOV equation is a set of equations that describe the structure of a static, spherically symmetric body\cite{Oppenheimer}\cite{Tolman}. It is the analogue of Newton's hydrostatic equilibrium equation that describes the equilibrium properties of a spherically symmetric body\cite{Oppenheimer}. It completely describes the structure of a spherically symmetric body when coupled with the EoS\cite{Tolman}.
	%
	%
	%
	\section{Lambda hyperon equation of state}
\begin{figure}[h]
		\includegraphics[width=18pc]{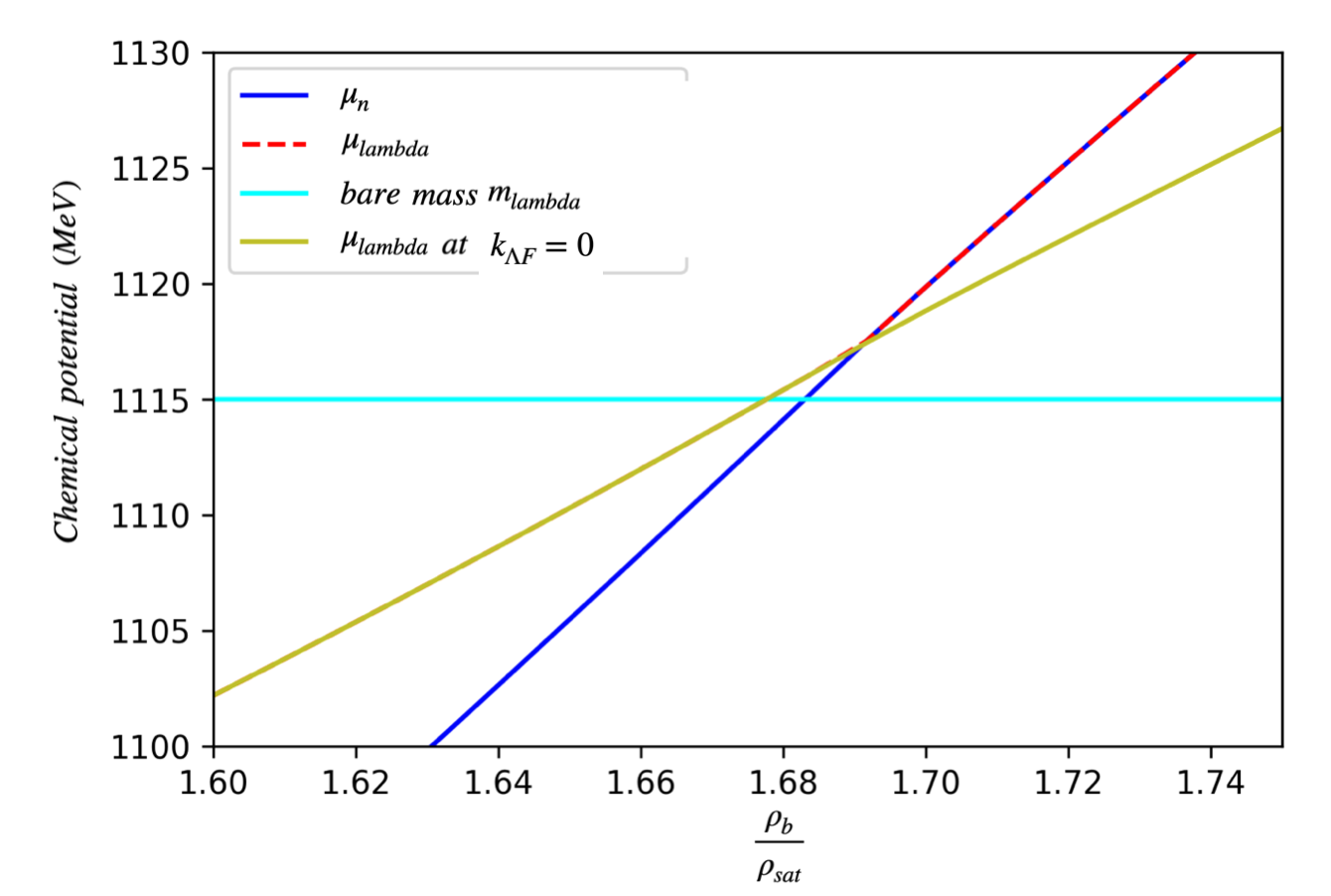}\hspace{2pc}%
		\begin{minipage}[b]{15pc}\caption{\label{fig:10.4}
				Contributions to $\mu_{\Lambda}$ (\ref{Lambda}), showing the density at which the lambda hyperon state are being populated in the QHD-I parameter set. 
			}
		\end{minipage}
	\end{figure}

\begin{figure}[h]
		\begin{minipage}{16pc}
			\centering
			\includegraphics[width=16pc]{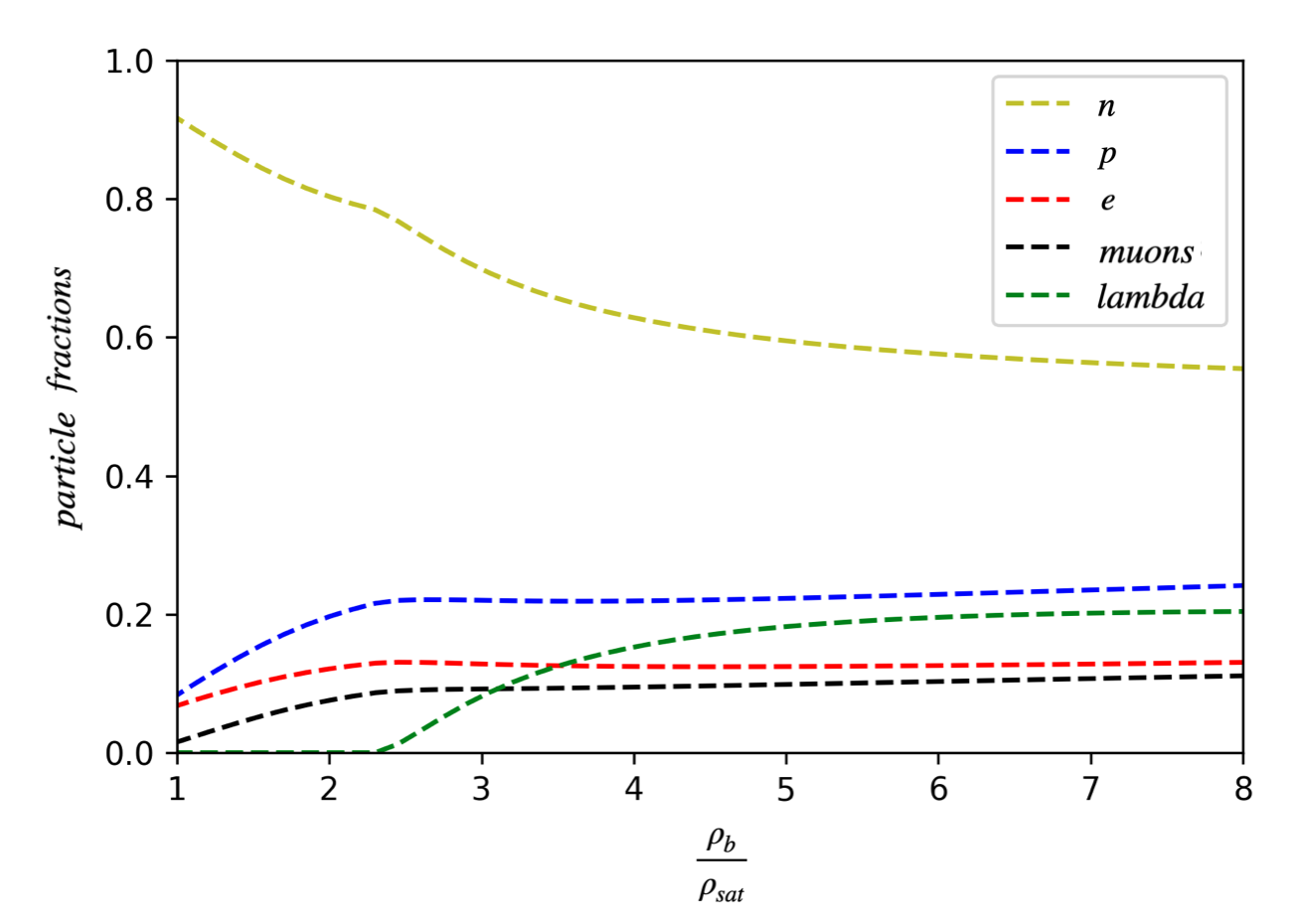}\\
			\caption{Particle fractions vs $\frac{\rho_b}{\rho_{sat}}$ (expressed as a multiple of the saturation density) for neutrons ($n$), electrons ($e$), protons ($p$), muons, and the lambda hyperon. The lambda hyperon appears at a density above the $\rho_{sat}$ of  ${1.69 \rho_{sat}}$ for NL3 parameter set for a system in general equilibrium with the lambda hyperon included. }
			\label{fig:10.5}
		\end{minipage}\hspace{3pc}%
		\begin{minipage}{16pc}
			\includegraphics[width=17pc]{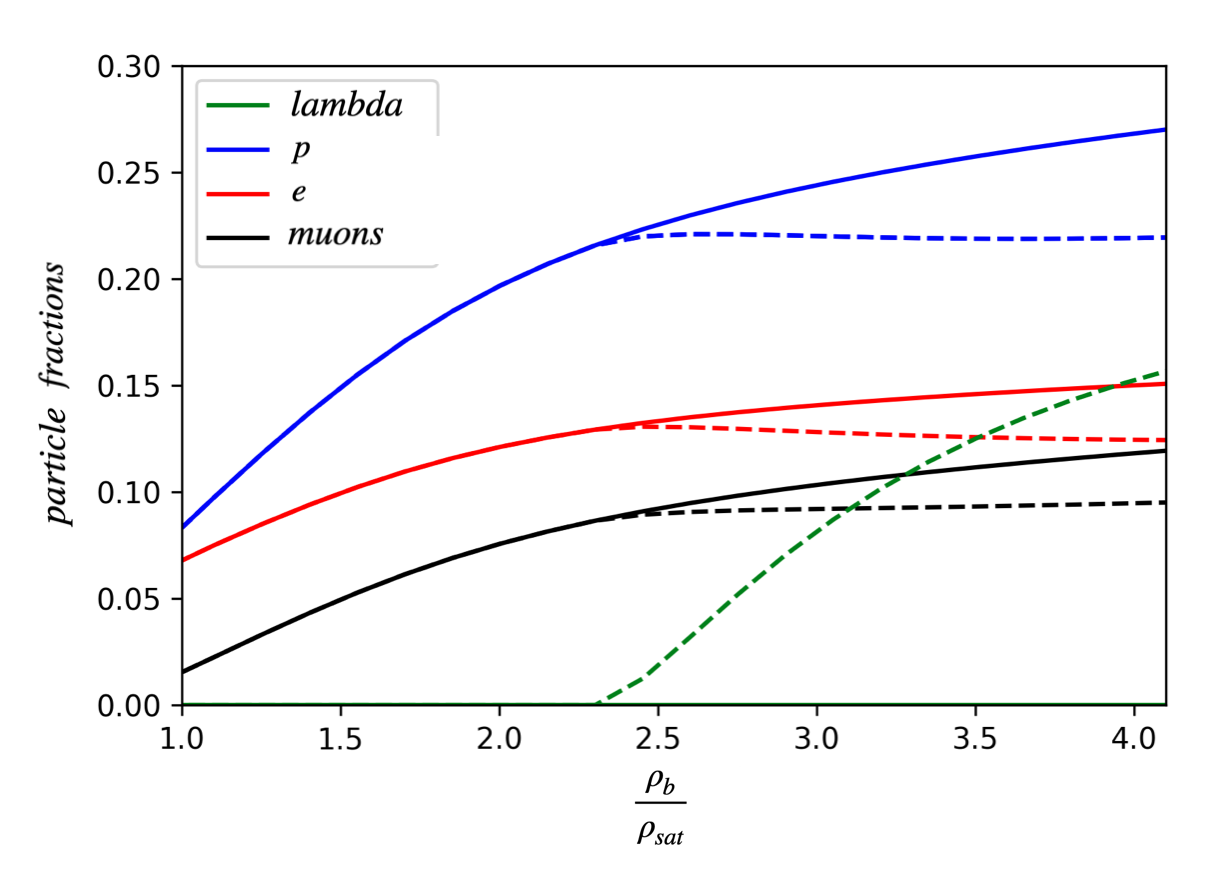}\\
			\caption{A close look at Fig \ref{fig:10.5} with the area of interest showing only the particle fractions for protons, electrons and muons before hyperons appear in the system and after adding hyperons for NL3. The solid lines represent general equilibrated matter and the dashed lines represents general equilibrated matter with the lambda ($\Lambda$) hyperon included.}
			\label{fig:10.6}
		\end{minipage} 
	\end{figure}

	The lambda hyperon hyperon is the least massive hyperon and will be the first hyperon state to be populated. It has zero charge, a bare mass of $1115$ MeV, only slightly larger than the nucleon mass ($939$ MeV),  and thus ideal to be included as a first approximation to study the effect hyperons have on the neutron star EoS \cite{Glendenning}.\\
	\\  
	The lambda hyperon state will be populated when two nucleons combine to produce a hyperon,
	\begin{eqnarray}
		N + N \longrightarrow N + \Lambda + K,
	\end{eqnarray}
	which also produces a kaon ($K$), which can decay to photons or transitions to a condensed state \cite{Glendenning}.  Here we will not consider $K$ interactions in our model.\\
	\\
	From Eq.\ \ref{Lgrngn} the chemical potential of the lambda hyperon is
	\begin{eqnarray}
		\mu_{\Lambda} &=& \sqrt{k_{\Lambda}^{2} + m^{*2}_{\Lambda}} + g_{v}V_{0} \label{Lambda},
	\end{eqnarray}
	where $m^*=M_{\Lambda}-g_{s}\phi_0$ is reduced mass of the lambda with $\phi_0$ the ground state expectation value of the scalar meson field. In order for the lambda hyperon ($\Lambda$) states to be populated the chemical potential of the nucleon/s ($N$) have to increase above the zero momentum ($k_{\Lambda} = 0$) chemical potential of the lambda hyperon (not the bare mass of the lambda due to the contributions of the meson fields). \\
	\\
	The appearance of the lambda hyperons are illustrated in figure \ref{fig:10.4}.  There the different contributions to Eq. \ref{Lambda} are compared for the QHD-I parameter set, where the lambda hyperons will start populating the system at a density of $1.69 \rho_{sat}$, illustrating that hyperons are a high density effect and not observed in nuclear matter under normal circumstances.  For NL3 parameters, the hyperons enter the system at $2.31 \rho_{sat}$.\\
	\\
	%
	Fig. \ref{fig:10.5} shows a plot of the particle fractions comparing the fractions of various baryons and leptons to the total baryon density of the system for NL3 parameter set. As a high energy neutral particle, the appearance of the lambda hyperons significantly reduces the neutron fraction.  Fig. \ref{fig:10.6} compares the particle fractions with and without hyperons in the system, which shows that the conversion of neutrons to lambda hyperons stabilises the charged particle fractions, which from the appearance of the lambda hyperons are also constant.  \\
	%
	%
	\section{Discussion}
\begin{figure}[h]
		\begin{minipage}{16pc}
			\centering
			\includegraphics[width=16pc]{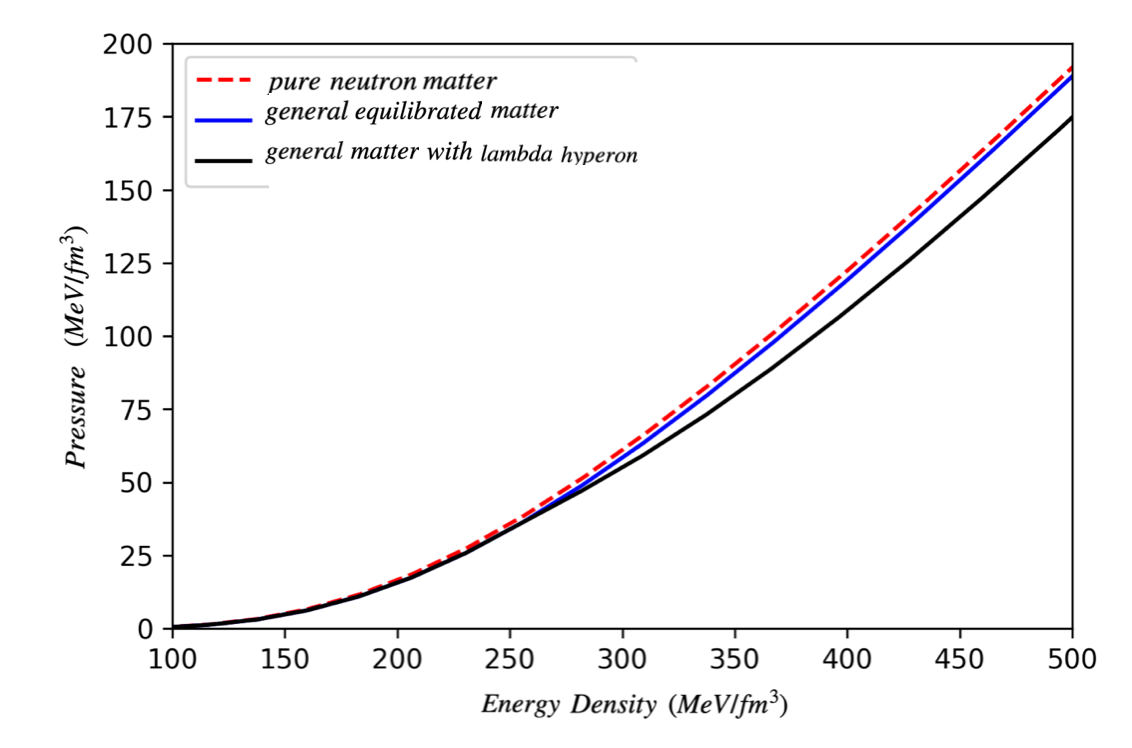}
			\caption{Pressure as function of energy density (EoS) for nuclear matter systems with the indicated equilibrium conditions in the QHD-I parameter set. The range of energy density roughly spans the baryon density range of $(0.1$-$8.0)  \rho_{sat}$.}
			\label{fig:EoSQ}
		\end{minipage}\hspace{3pc}%
		\begin{minipage}{16pc}
			\includegraphics[width=17pc]{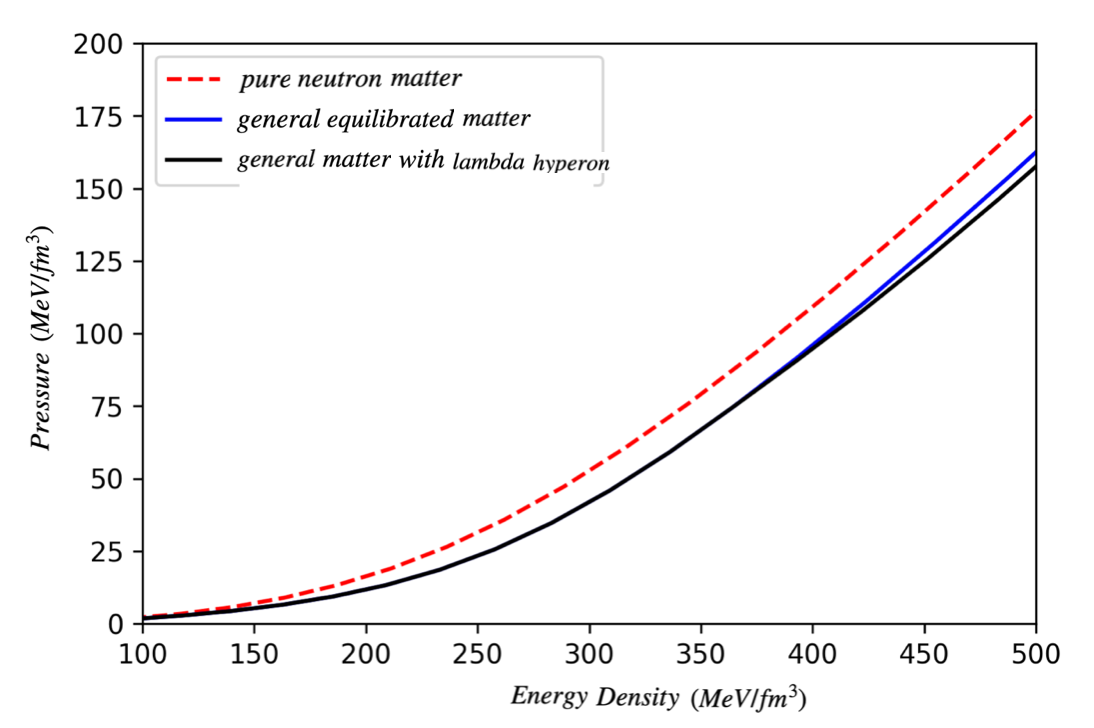}
			\caption{Pressure as function of energy density (EoS) for nuclear matter systems with the indicated equilibrium conditions in the NL3 parameter set. The range of energy density roughly spans the baryon density range of $(0.1$-$8.0)  \rho_{sat}$.}
			\label{fig:EoSN}
		\end{minipage} 
	\end{figure}

\begin{figure}[h]
		\begin{minipage}{16pc}
			\centering
			\includegraphics[width=16pc]{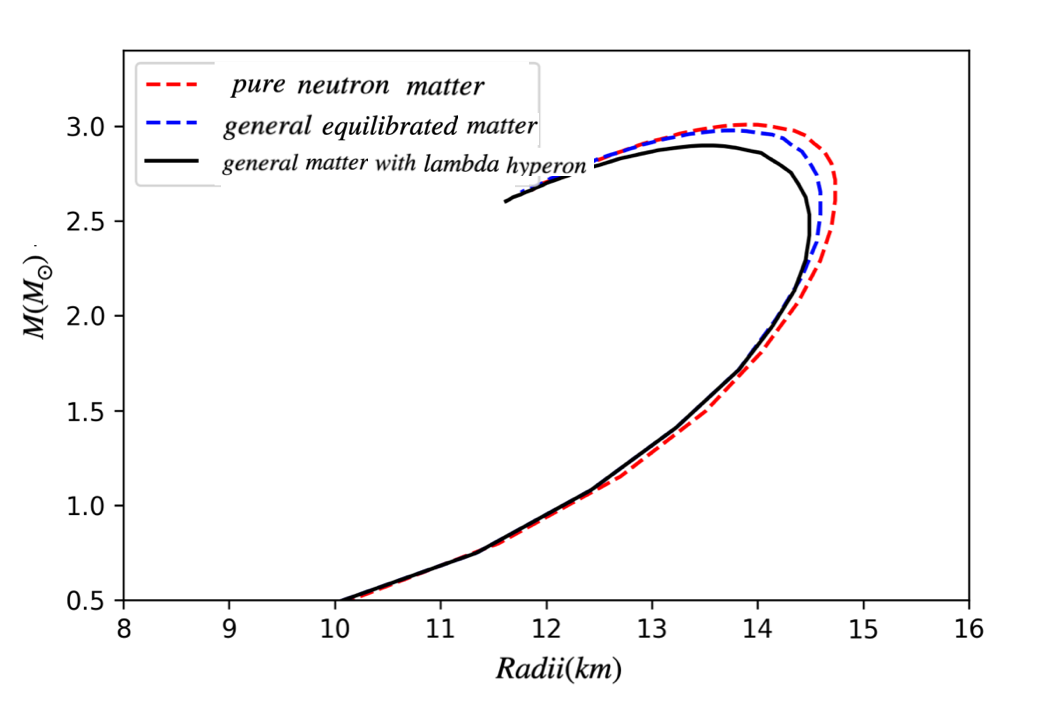}
			\caption{The mass-radius relationship for pure neutrons, general equilibrated matter and general equilibrated with the lambda hyperon included for QHD-I.}
			\label{fig:MRQ}
		\end{minipage}\hspace{3pc}%
		\begin{minipage}{16pc}
			\includegraphics[width=17pc]{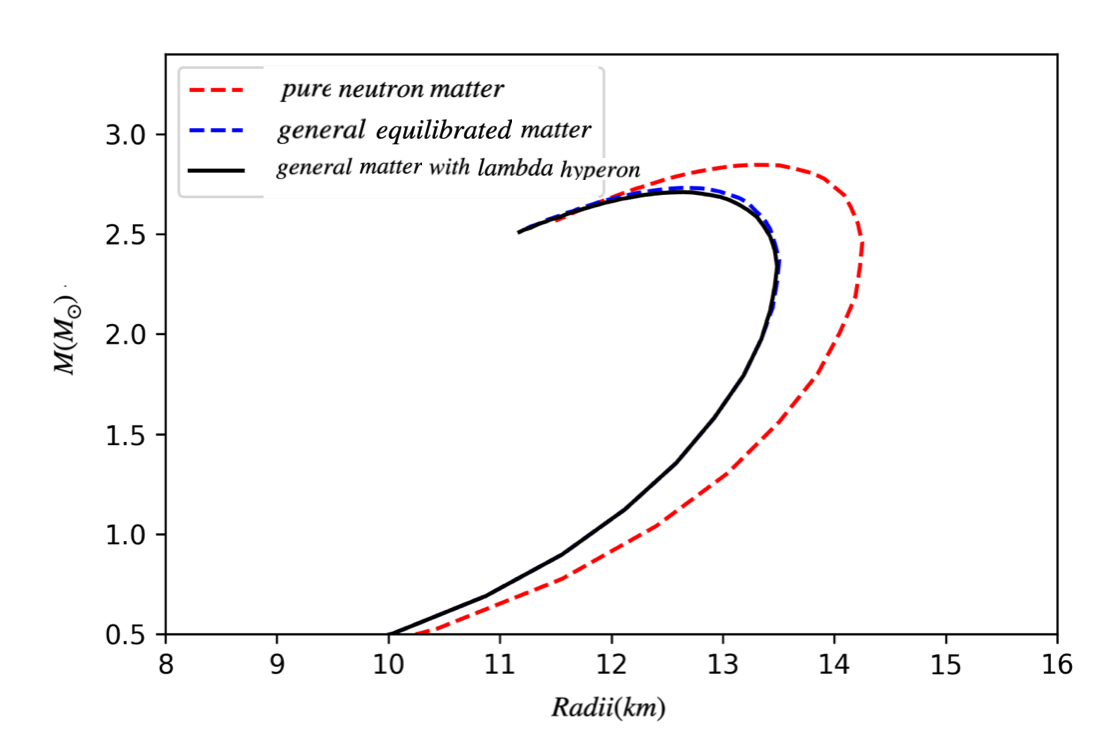}
			\caption{The mass-radius relationship for pure neutrons, general equilibrated matter and general equilibrated with the lambda hyperon included for NL3.}
			\label{fig:MRN}
		\end{minipage} 
	\end{figure}

	The addition of the  neutral lambda hyperons as a degree of freedom to neutron star EoS
	provides a mechanism that softens the EoS since the added 
	degree of freedom is clearly preferred as shown in Fig.\ \ref{fig:10.5}. The softening of the EoS is also clearly illustrated by direct comparison to other EoS, as shown in figures \ref{fig:EoSQ} and \ref{fig:EoSN}, where the delayed increase in pressure is obvious. This consequently results in the reduction of the maximum masses as shown in the mass-radius (M-R) plot figures \ref{fig:MRQ} and \ref{fig:MRN}.\\
	\\
	However, comparing the behaviour of the NL3 parameter set to that of QHD-1 it is clear that both the EoS and the corresponding M-R relationships are sensitive to the densities at which hyperons appear in the system: due to the comparatively higher density appears of the hyperons in NL3 seems to drastically reduce the impact on the M-R relationship. As such, at first glance, the severity of the hyperon puzzle can be negated by the choice of parameter set to describe nuclear matter with hyperons.\\
	\\
	However, what is of course needed is more observational and experimental results that can indicate the high density behaviour of nuclear matter that can improve our understanding of the appearance of the lambda hyperon in equilibrated nuclear matter systems. 
	Fortunately observational data from massive new radio-telescopes like the Square Kilometre Array (SKA) will provide observations that can be supported and evolve theoretical models of nuclear matter. 
	\section{Conclusion and future work}
	While the sole inclusion of the lambda hyperon to the description of dense nuclear matter found in neutron star is not unrealistic, a more complete approach would be to include the full baryon octet in the description of nuclear matter inside a neutron star. The expectation is that the addition of more hyperons should further soften the equation of state \cite{Bombaci}. However, based on our finding the degree of softening will be model dependant. This will be the focus of our future work. 
		

\section*{References}
	\begin{thebibliography}{9}
		\bibitem{Serot} Serot, B D and Walecka, J D
		{1992} 
		{\it Recent Progress in Many-Body Theories}
		vol 16,
		ed. J W Negele and E Vogt (Plenum Press: New York - London)
		{p 49}
		
		
		\bibitem{Glendenning}Glendenning, N K 
		{2000}
		{\it Compact Stars: Nuclear Physics, Particle Physics, and General Relativity}
		{(Springer: New York)}
		
		\bibitem{Prakash} Prakash, M and Bombaci, I and Prakash, M and Ellis, P J and Lattimer, J M and Knorren, R,
		{1997},
		{\it Composition and structure of protoneutron stars},
		{\it Physics Reports},
		{\bf 280},
		{1},
		{1--77},
		{Elsevier}
		
		\bibitem{Haensel}Haensel P, Potekhin A  and Yakovlev D G 2007 {\it Neutron stars 1: Equation of state and structure} 
		{Springer: New York }
		
		\bibitem{Vidana}Vidana, I 2016 
		{\it J. Phys.: Conf. Series} {\bf 668}  {012031} 
		
		\bibitem{Demorest}Demorest, P B and Pennucci, T and Ransom, S M, Roberts M S E and Hessels J W T
		{2010}
		{\it Nature},
		{\bf 467},
		1081
		
		\bibitem{Antoniadis}Antoniadis, J 
		\textit{et al.}
		{2013}
		{\it Science},
		{\bf 340},
		{1233232}
		
		
		
		\bibitem{Bombaci}Bombaci, I 2017 
		{\it JPS Conf.Proc}
		\textbf{17} 101002
		
		\bibitem{Logoeta}Logoteta, D 2021 
		{\it Universe} {\bf 7}
		{408}
		\bibitem{NL3} Lalazissis G A,  K\"{o}nig J, and  Ring P  1997  \textit{Phys. Rev. C} \textbf{55}   540
		\bibitem{Lattimer} Lattimer, J M and Prakash, M 2004
		{\it Science}
		{\bf 304}
		536
		\bibitem{Oppenheimer}Oppenheimer, J R and Volkoff, G M 1939 {\it Physical Review}
			{\bf 55}
			{374}
			
		\bibitem{Tolman}Tolman, R C 1939 {\it Physical Review}
			{\bf 55}
			{364}
			
		
		
		
		
	\end{thebibliography}
\end{document}